\documentclass[aps,prb,reprint]{revtex4-2}
\usepackage{blindtext}
\usepackage{graphicx}
\usepackage{amsmath}
\usepackage{upgreek}
\usepackage[english]{babel}

\usepackage{braket}
\usepackage{hyperref}
\hypersetup{colorlinks=true,linkcolor=blue, filecolor=blue, urlcolor=blue, citecolor=blue}
\usepackage{color}
\definecolor{darkgreen}{rgb}{0.01, 0.75, 0.24}

\definecolor{darkred}{rgb}{0.72, 0.11, 0.17}

\begin{document}
\title{Quantum electrometry of non-volatile space charges in diamond}
\author{R. M. Goldblatt}
\author{N. Dontschuk}
\author{D. J. McCloskey}
\author{A. M. Martin}
\author{A. A. Wood}
\email{alexander.wood@unimelb.edu.au}
\affiliation{School of Physics, University of Melbourne, Parkville Victoria 3010, Australia}
\date{\today}

\begin{abstract}
The microscopic electric environment surrounding a spin defect in a wide-bandgap semiconductor plays a determining role in the spin coherence and charge stability of a given qubit~\cite{wolfowicz_quantum_2021} and has an equally important role in defining the electrical properties of the host material~\cite{fedyanin_optoelectronics_2021, zhang_perspective_2023}. Here, we use electrometry of quantum defects embedded within a diamond to observe stable, micron-scale space charge distributions formed from trapped photogenerated charges. These space charges grow under optical illumination in the presence of an applied electric field, eventually screening the applied electric field entirely over a spatial extent of tens of microns due to charge carrier drift and capture. Our measurements suggest that these space charge fields originate from widely-dispersed spatial configurations of nitrogen charges. Our results have important consequences for electrometry and photoelectric detection using qubits in wide-bandgap semiconductors.
\end{abstract}
     
\maketitle

Spin qubits are never created in perfect isolation within a solid state lattice, and a range of nearby impurities represent sources of magnetic or electrical noise that can perturb a central spin during quantum measurement.  Charge-state transfer processes between the qubit and the environment result in imperfect state initialisation and shot-to-shot variations of local electromagnetic noise~\cite{dolde_nanoscale_2014, ji_correlated_2024, delord_correlated_2024, pieplow_quantum_2024}. The addition of external electric fields, whether for the purposes of sensing~\cite{dolde_electric-field_2011}, charge state identification~\cite{dhomkar_-demand_2018} and control~\cite{schreyvogel_active_2015,rieger_fast_2023}, photoelectric detection~\cite{bourgeois_photoelectric_2015, siyushev_photoelectrical_2019,niethammer_coherent_2019,bourgeois_photoelectric_2020,zheng_electrical-readout_2022}, or tuning photophysical properties for improved optical readout~\cite{hanlon_spin--charge_2021,hanlon_enhancement_2023}, has significant though poorly understood effects on the spin-charge environment of a quantum defect. In particular, the dominance of deep trap levels and lack of room temperature carriers in a wide-bandgap material like diamond can result in the formation of stable, macroscopic regions of charged impurities -- `space charges' -- which screen applied electric fields~\cite{bassett_electrical_2011, lozovoi_probing_2020,oberg_solution_2020}, complicate carrier transport dynamics~\cite{forneris_mapping_2018, wood_3d-mapping_2024} and degrade the performance of radiation detectors~\cite{naaranoja_space_2019, shimaoka_recent_2022}. 

In this work, we show that charge photogeneration, transport and capture by other defects~\cite{jayakumar_optical_2016} in the presence of an external electric field results in the formation of large, stable space charges that can totally screen the applied electric field over spatial regions of tens of microns. We use quantum sensors embedded in the material -- nitrogen vacancy (NV) centres~\cite{doherty_nitrogen-vacancy_2013} -- to probe the electric fields created by charged impurities that capture photoelectrons to dynamically form space charges that oppose the applied electric field. We provide strong evidence to suggest the abundant substitutional nitrogen atoms~\cite{ashfold_nitrogen_2020} are responsible for the formation of space charge potentials. These space charges in turn have profound implications for the transport of charges, resulting in counterintuitive carrier drift and intricate spatial dynamics~\cite{wood_3d-mapping_2024}. Our work will have important consequences for diamond quantum electrometers, photoelectric detection of spin qubits in semiconductors and potentially the study of diamond and other wide-bandgap materials in power electronics~\cite{wort_diamond_2008, donato_diamond_2019, araujo_diamond_2021, scheller_quantum_2024}. 

The NV centre in diamond serves as an exemplar quantum spin system in a wide-bandgap material to investigate the interplay between external and internal electric fields. While many notable demonstrations have been reported using the room-temperature ground state~\cite{dolde_electric-field_2011, doherty_measuring_2014,chen_high-sensitivity_2017,michl_robust_2019} or low-temperature excited-state~\cite{tamarat_stark_2006, block_optically_2021, ji_correlated_2024, delord_correlated_2024} Stark shifts, the applications of the NV centre in diamond to room-temperature electrometry are already limited by the very weak ground state electric susceptibility parameters of $(d_\perp, d_\parallel) = (17, 0.3)$\,Hz\,V$^{-1}$\,cm$^{-1}$ for transverse and longitudinal electric fields~\cite{van_oort_electric-field-induced_1990}. The presence of a dynamically reconfigurable charge environment that results in space charge effects serves to screen the applied electric field at the sensor, further reducing sensitivity. Additionally, these charged defects result in an electric field response convolved with the sample-specific local charge environment of a given NV defect~\cite{mittiga_imaging_2018}, which is also affected by local crystal strain~\cite{kolbl_determination_2019}. 

Extensive work has examined the effect of electrical noise~\cite{kim_decoherence_2015,jamonneau_competition_2016} and Debye screening~\cite{oberg_solution_2020} on NV electrometry~\cite{stacey_evidence_2019, bian_nanoscale_2021, barson_nanoscale_2021,zheng_coherence_2022} with a focus on the role of the nearby surface,  a choice motivated by the necessity of reducing the standoff distance between the sensing defect and an external target.  Nanodiamonds mounted to scanning cantilevers in atomic force microscopes have been used to image electric fields within other materials~\cite{huxter_imaging_2023}.  Other work has drawn attention to the significant \emph{internal} electric fields created inside diamond by both near-surface band bending~\cite{broadway_spatial_2018} and local engineering of defect charge states and coherence~\cite{zheng_coherence_2022,kobayashi_electrical_2020}. The presence of both external and impurity-mediated internal electric fields also affects the transport and capture of carriers between single defects~\cite{lozovoi_optical_2021, lozovoi_imaging_2022, lozovoi_detection_2023}, which have attracted interest as nodes for the distribution of quantum information~\cite{doherty_towards_2016,oberg_spin_2019}. While our work focuses on the NV centre in diamond, we point out that the methods and results in this work are equally applicable to defects in other wide-bandgap materials that exhibit Stark shifts, such as V$^-$ or VV$^0$ in SiC~\cite{de_las_casas_stark_2017,wolfowicz_electrometry_2018,ruhl_stark_2020,bathen_electrical_2019,wolfowicz_heterodyne_2019}. 

\begin{figure*}
	\centering
		\includegraphics[width = \textwidth]{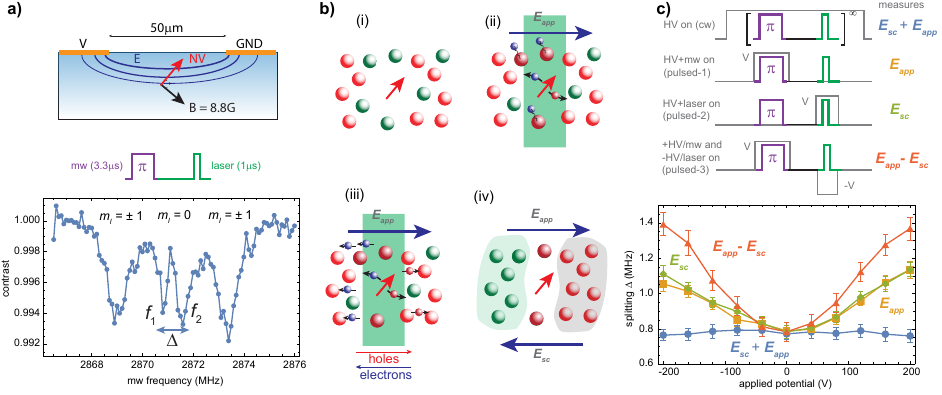}
	\caption{\textbf{Formation and measurement of space charges.} (a) top: schematic of the experiment: NV centres a few microns below the surface of the diamond measure the electric field generated by planar surface electrodes. An $8.3\,$G magnetic bias field $\mathbf{B}$ orthogonal to one NV axis confers sensitivity to the Stark shift.  Bottom: the electric field from both $E_\text{app}$ and nearby charges manifests in an ODMR measurement as a splitting of the central, $m_I = 0$ nuclear spin hyperfine feature. (b) Formation of a space charge field that opposes an applied electric field. In (i), an initially unpolarised charge environment in the presence of $E_\text{app}$ is illuminated by green light, generating free carriers from photoionisation and recombination (ii). The carriers drift under the applied field (iii) and are re-trapped, forming a charge distribution and space charge field $E_\text{sc}$ that cancels the applied field. (c) top, pulse sequences (cw, pulsed-1,2,3) to measure $E_\text{app}+E_\text{sc}$, $E_\text{app}$, $E_\text{sc}$ and $E_\text{app}-E_\text{sc}$, respectively. Extracting the splitting $\Delta f$ from Gaussian fits to the ODMR data (bottom) reveals a space charge field exactly equal and opposite to the applied field. Error bars represent fitted Gaussian centroid standard errors.}
	\label{fig:fig1}
\end{figure*}

Our experiment, depicted schematically in Fig. \ref{fig:fig1}(a) consists of a CVD diamond sample with (100)-faces, nitrogen and NV concentrations of approximately 1\,ppm and 0.01\,ppm respectively, and Cr/Au electrodes deposited onto the surface. A confocal microscope featuring a 0.8\,NA apochromatic microscope objective on a three-axis scanning piezo stage illuminates NV centres with green (532\,nm, 0-8\,mW), red (633\,nm, 0-5\,mW) or orange (594\,nm, 7\,$\upmu$W) light and measures fluorescence emission selectively from negatively-charged NV centres (691-720\,nm), while a nearby copper wire generates microwave fields for quantum state manipulation. Three current-carrying coil assemblies provide a DC magnetic field of up to 10\,G in any direction, and a fast high-voltage amplifier is used to apply bias voltages to the electrodes. All our experiments are conducted at room temperature.

Electrometry using the ground state spin of the NV centre requires eigenstates susceptible to the Stark shift. The NV ground state spin Hamiltonian in the presence of an electric field $\mathbf{E} = (E_x, E_y, E_z)$ and magnetic field $\mathbf{B}$ is given by 
\begin{align}
H & = (D + d_\parallel E_z)S_z^2 + \gamma \mathbf{B}\cdot\mathbf{S} \\
 & - d_\perp\left[ E_x (S_x^2 - S_y^2) + E_y (S_x S_y + S_y S_x)\right],
\label{eq:ham}
\end{align}
with $D = 2.87$\,GHz the ground-state zero field splitting, $\gamma = 2.8\,$MHz\,G$^{-1}$ the electron gyromagnetic ratio and $\mathbf{S} = (S_x, S_y, S_z)$ the NV spin, with the $z$-axis taken to be the N-V axis and the $x$-axis in one of the C-N bond planes. We use an 8.3\,G magnetic field oriented parallel to the $x$-axis (orthogonal to the NV axis) to generate Stark sensitive eigenstates $|\pm\rangle$ from a superposition of NV $|m_S = \pm1\rangle$ states~\cite{michl_robust_2019}. The presence of the unpolarised intrinsic $^{14}$N $I = 1$ nuclear spin of the host NV results in six partially overlapping features in an optically-detected magnetic resonance (ODMR) experiment, as shown in Fig. \ref{fig:fig1}(b): two sets of degenerate electron spin transitions corresponding to the $m_I = \pm1$ states and a central, split feature corresponding to the Stark-sensitive $|m_S = 0, m_I = 0\rangle \rightarrow |\pm, m_I = 0\rangle$ transitions, which we call $f_-$ and $f_+$ in ascending frequency order. 
 
Neglecting the much weaker axial electric susceptibility $d_\parallel$\cite{michl_robust_2019}, the splitting $\Delta f = f_+ - f_-$ as a function of an applied electric field with component $E_\perp$ orthogonal to the NV axis is
\begin{align}
\Delta f(E_\perp) = 2\sqrt{R^2+\Lambda^2-R \Lambda \cos\,\phi_E}+\Delta f_0,
\label{eq:f12}
\end{align}
where $\Delta f_0$ is the splitting in the absence of an electric field, $R = d_\perp E_\perp$, $\Lambda=(\gamma_e B)^2/2D$ and $\phi_E$ the azimuthal angle $E_\perp = \sqrt{E_x^2+E_y^2}$ makes to the applied magnetic field ($x$-axis) in the NV frame. An electric field $\mathbf{E}_\text{app}$ generated from the planar electrodes on the diamond surface is assumed to be predominantly lateral, and thus appears as $E_\perp = \eta |\mathbf{E}_\text{app}|$ with $\phi_E = 69.2^\circ$ and $\eta\approx 0.89$ in Eq. \ref{eq:f12}. The splitting in the absence of an applied electric field arises due to the transverse magnetic field~\cite{doherty_measuring_2014} and local charge environment of the NV centres~\cite{mittiga_imaging_2018}, and is well explained by an estimated $0.3\,$ppm N concentration. The presence of the transverse magnetic field results in a slightly nonlinear response of $\Delta f$ at electric fields below $10\,$kV\,cm$^{-1}$, but beyond this the splitting is well approximated by $\Delta f\approx 2d_\perp\eta |\mathbf{E}_\text{app}|$ and we use this henceforth to estimate the electric field strength. 

{\bf Space charge formation.} A space charge is formed during a measurement of the applied electric field $E_\text{app}$ via the mechanism outlined in Fig.\ref{fig:fig1}(b). Green illumination of the NV centre and its environs (i) generates electrons and holes (ii) through ionisation of traps such as N$^0$ (photoionisation threshold of 2.2\,eV~\cite{bourgeois_enhanced_2017}) or two-photon charge cycling of the NV itself~\cite{aslam_photo-induced_2013}. These liberated carriers drift according to the applied field (iii) before being trapped once again, with N$^+$ +e$^-\rightarrow$N$^0$ being most prominent given the lack of coulombic attraction between charge carriers and neutral defects. The net migration of charge due to the applied field results in a charge-polarised spatial distribution (iv), creating a space-charge field $E_\text{sc}$ that grows with subsequent repetitions of the experiment. This cumulative growth process is crucial, and occurs because the short green readout pulses in typical pulsed ODMR measurements are insufficiently powerful or long to reinitialise the NV charge environment. We return to the time and power dependence of space charge distributions later.

Fig. \ref{fig:fig1}(c) shows the results of several variations on pulsed ODMR, which alternate the times when lasers ($t_l = 1.5\,\upmu$s, 2.5\,mW), microwave pulses ($t_\pi = 3.3\,\upmu$s) and the electrode potentials are simultaneously on. In `cw' electrometry, the electric field is applied continuously, while the laser and microwave fields are pulsed alternately in a sensing configuration that emulates the detection of an uncontrolled DC electric field. Here, we observe no change in the splitting $\Delta f(E)$, as the simultaneous laser and electric field generates an $E_\text{sc}$ that opposes the $E_\text{app}$, so that the electric field sampled by the microwave pulse is $E_\text{sc} + E_\text{app} = 0$. Similar effects were observed with surface-induced screening in Ref. \cite{barson_nanoscale_2021}. Pulsing the electric field simultaneously with the microwaves (`pulsed-1') but not during the laser pulse prevents the accumulation of a space charge so that the microwaves only sample $E_\text{app}$, while pulsing the HV and laser simultaneously (`pulsed-2') creates only a space charge field $E_\text{sc}$ that is probed with the microwaves, noting that the applied field $E_\text{app}$ is off at this time. 

Finally, we invert the polarity of the potential applied during the laser pulse to be opposite that when the microwave pulse is applied (`pulsed-3'), which measures $E_\text{app} - E_\text{sc}$, and we consequently observe nearly twice the splitting for a given applied voltage. The nonlinear response to weak electric fields is evident in the data shown in Fig. \ref{fig:fig1}(c), however the nonlinear response to the largest applied total fields in pulsed-3 is presently unexplained. Nevertheless, our results demonstrate that $E_\text{sc} + E_\text{app} = 0$ when measured together (cw) and $E_\text{app} - E_\text{sc} \approx 2 E_\text{app}$ (pulsed-3), while we also see independently that $|E_\text{sc}|=|E_\text{app}|$ (pulsed-1,2). NV electrometry is insensitive to the sign of the electric field~\cite{dolde_electric-field_2011, doherty_measuring_2014, michl_robust_2019}, but we see clear evidence that the space-charge field acts to cancel the applied field. 

{\bf Optical power dependence.} We now proceed to perform a detailed characterisation of the formation process and hence origin of the space charge fields. As the most abundant impurity in essentially every diamond sample (including our own) and possessing stable neutral and positive charge states at room temperature~\cite{ashfold_nitrogen_2020}, substitutional N$_S$ is the most likely candidate to facilitate the formation of a space charge. We note that a negative charge state is not necessary for a space charge, which can be generated by a spatial distribution of neutral and positively-charged impurities. 

\begin{figure}
	\centering
		\includegraphics[width = \columnwidth]{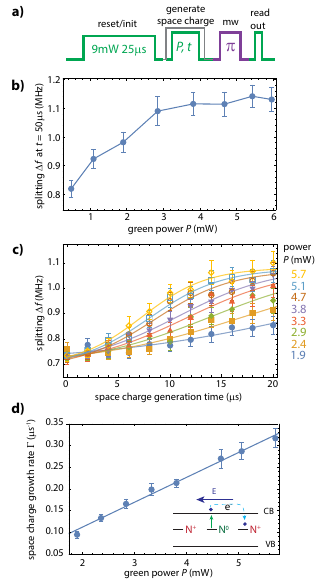}
	\caption{\textbf{Optical power dependence of space charge generation.} (a) A strong reset pulse (25$\,\upmu$s, 9\,mW) breaks the cumulative generation of space charges, allowing the power and time dependence to be probed with a variable duration space charge generation pulse protocol. Grey denotes application of a +200\,V potential to the left-hand electrode, experiments are conducted 10\,$\upmu$m from the electrode edge. (b) A long generation pulse (50\,$\upmu$s) at various powers shows that the space charge field saturates to a constant value, 0.389(22)\,MHz or approximately 20\,kV\,cm$^{-1}$. (c) The time dependence of the space charge formation has a sigmoidal shape, well described by a logistic fit (solid lines).  (d) The generation rate in Eq. \ref{eq:sig} depends linearly on the optical power, consistent with nitrogen photoionisation and electron capture (inset). Error bars in (b,c) are standard error in fitted Gaussian centroid splittings, in (d) error bars in $\Gamma$ parameters from least-squares fits of Eq. \ref{eq:sig} to data in (c).}
	\label{fig:fig2}
\end{figure}

We first assess the role of cumulative effects in the formation of a space charge. The short (1.5\,$\upmu$s), 2.5\,mW laser pulse used in the `cw' ODMR electrometry sequence for spin initialisation in Fig. \ref{fig:fig1} is responsible for the generation of $E_\text{sc}$, though a single iteration alone is insufficient to generate a space charge that totally screens $E_\text{app}$. Without any active resetting of the charge environment, the space charge grows cumulatively between experimental repetitions. We therefore examine in Fig. \ref{fig:fig2}(a) the time and power dependence of space charge formation with a strong reset pulse that breaks cumulative growth. The pulse sequence begins with a strong 532\,nm laser pulse (9\,mW, 25$\upmu$s) applied in the absence of an electric field. A variable-power, variable-time green laser pulse applied simultaneously with a 200\,V electrode potential subsequently generates a space charge, which is then measured with a pulsed ODMR sequence (pulsed-2). The strong green pulse prevents cumulative growth of the space charge between experimental repetitions, and allows the single-shot time and power dependence of the space charge growth to be assessed. The results are shown in Fig. \ref{fig:fig2}(b), and reveal a gradual onset of space-charge growth before saturation in a sigmoidal shape. 

The space-charge splitting vs. time data is well approximated by a logistic curve, 
\begin{equation}
\Delta f (t) = A\left(\frac{e^{\Gamma t}}{D+e^{\Gamma t}}-\frac{1}{D}\right),
\label{eq:sig}
\end{equation}
which represents the solution to the differential equation $\Delta f'(t) = \Gamma \Delta f(t)(A -\Delta f(t))$ that describes exponential growth at a rate $\Gamma$ followed by saturation to an asymptote $A$. The dimensionless parameter $D$ sets the initial rise characteristic, and was approximately 10 for all the data. Before fitting with Eq. \ref{eq:sig}, we confirmed the space charge splitting saturates to a power-independent asymptote $A$ by generating a space charge with a 50\,$\upmu$s green pulse at various powers, the data in Fig. \ref{fig:fig2}(c) show the space-charge splitting asymptoting to a final value of $A = 0.389(21)\,$MHz. Fitting to the data and extracting the exponential growth parameter $\Gamma$, we observe a linear increase in space charge generation rate with optical power,  shown in Fig. \ref{fig:fig2}(d).  The linear dependence is characteristic of a process driven by single photon mediated ionisation, most likely of the abundant nitrogen impurities.  

{\bf Spatial scale measurements.} The results in the previous section implicate nitrogen as the trapping centre responsible for the formation of a space charge. However, the measurements do not reveal the specific spatial configuration of the charged nitrogen distribution. To determine if the space charge is nanoscopic (in the immediate vicinity of a given NV centre), microscopic (within the extent of the laser spot) or macrosopic (larger than the laser spot), we added a second 532\,nm laser beam into the microscope, independently controlled with a scanning galvanometer mirror system~\cite{rovny_nanoscale_2022}. Here, the position of the microscope objective (controlled by a three-axis piezo scanner) sets the position of one beam (beam-1) while the galvo system controls the relative position of a second (beam-2). 

\begin{figure*}
	\centering
		\includegraphics[width = \textwidth]{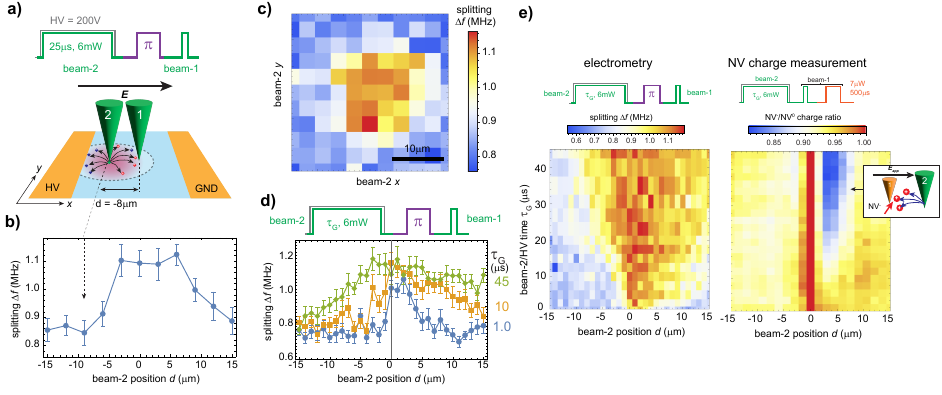}
	\caption{\textbf{Electrometry of spatially extended charge distributions.} (a) A dual beam confocal microscope enables generation of a space charge and ODMR measurement on spin relaxation timescales without optical perturbation of the probe charge environment. The space charge electric field profile generated by a 25\,$\upmu$s, 6\,mW beam-2 (movable) pulse at 200\,V potential is probed with beam-1 (fixed), revealing an almost 20\,$\upmu$m region of polarised charge trapping in one (b) and two (c) spatial directions. (d) Varying the time of the space-charge generation pulse $\tau_G$ shows that the space charge profile develops anisotropically, first strongly biased to the left (consistent with electrons from beam-2 traveling towards the probe region) before equilibrating at longer $\tau_G$. Wherever the space charge field is high, the total electric field $E_\text{app}+E_\text{sc}$ is low. (e) We compare the electron-driven spatiotemporal dynamics of space charge formation (left), to that of hole generation and capture by NV centres (right). Using a charge-sensitive orange readout pulse, we can trace the trajectories of holes generated via NV charge cycling, and observe that holes are captured by NV centres in regions of low total electric field strength, appearing to flow against the direction of the applied electric field (inset).}
	\label{fig:fig3}
\end{figure*}

The experimental sequence proceeds first with a 6\,mW, $\tau_G = 25\,\upmu$s beam-2 pulse and simultaneous application of a 200\,V potential to the electrodes to generate a space charge at a given position of beam-2. The generation of a space charge is followed immediately by a 3.3\,$\upmu$s microwave $\pi$-pulse and then by a 1.5\,$\upmu$s, 2.2\,mW beam-1 readout pulse, and is repeated over 20\,minutes while scanning the microwave frequency and detecting the NV PL in a pulsed-ODMR measurement as before. The space charge profile created by the variable-position beam-2 is probed by the fixed beam-1, as described in Fig. \ref{fig:fig3}(a) and similar to the scheme outlined in Refs. \cite{gardill_probing_2021, lozovoi_imaging_2022} for charge state interconversion. The addition of the second independent beam precludes the need to physically reposition the laser in between optical preparation of the probe region, generation of a space charge and measurement, which needs to be kept below $T_1 = 2\,$ms for ODMR to be conducted.

We deliberately forgo the strong green initialisation step discussed previously to harness the cumulative effects of space charge formation. Without well-defined initialisation using a strong green laser for each repetition, the short readout green laser pulse is insufficient to adequately reset the charge environment. Therefore, a number of experimental repeats with a short space charge generation pulse $\tau_G$ has an almost equivalent effect to a single long generation pulse with resetting, and results in a steady-state space charge forming quickly after the first measurement (typically within 100-200 repetitions, of order milliseconds). Subsequent repeated measurements probe this steady-state charge distribution, building up the necessary averaged photon counts required to extract a high signal-to-noise ratio ODMR signal. These cumulative effects serve to amplify the effects of optically-induced space charges and improve measurement statistics without altering the underlying physics.  

We now present results which reveal the macroscopic spatial extent of the space charge generated by optical illumination. Fig. \ref{fig:fig3}(b) shows the ODMR splitting $\Delta f$ as a function of beam-2 position along the $x$ (inter-electrode) axis, following the protocol outlined in Fig. \ref{fig:fig3}(a), for a 25\,$\upmu$s, 6\,mW space charge generation pulse under an applied potential of 200\,V. We observe an almost symmetric region of saturated space charge extending over nearly 25\,$\upmu$m, which serves as clear evidence that optically generated charges from the localised laser spot drift under the applied electric field for significant distances before being captured to generate an opposing space charge field. Our technique is readily extended into two dimensions by also scanning the $y$-position of beam-2, allowing for a comprehensive evaluation of the space charge profile, as depicted in Fig. \ref{fig:fig3}(c). 

The roughly symmetrical shape of the space charge region in Fig. \ref{fig:fig3}(c) is somewhat surprising given the expectation carriers will drift according to the applied electric field, which should direct negative charges to the left in the images. The key to understanding this observation lies in the temporal dynamics of the space charge growth, as we now show. We modify the protocol in Fig. \ref{fig:fig3}(a) so that the space charge generation pulse duration is varied, and probe the $x$-position dependence at different times. Compiled in Fig. \ref{fig:fig3}(d,e) for $\tau_G$ ranging from $1$-$40\,\upmu$s, we see first that a narrow space charge region accumulates to the left of beam-2 (which appears as the right in Fig. \ref{fig:fig3}(d), see inset) before broadening significantly and then, eventually, spilling over to the opposite side of beam-2 and approaching the symmetric shape observed in Fig. \ref{fig:fig3}(b,c). 

{\bf Charge transport.} The spatiotemporal dynamics are a consequence of carrier transport and trapping that is dictated by the total electric field, which is zero everywhere the space charge field is saturated. As a particularly striking demonstration of this, we study the charge interconversion of the NV centres due to hole capture in the presence of the evolving space charge field, using the pulse sequence outlined in Fig. \ref{fig:fig3}(e) which removes the microwave pulse and measures the NV charge state using a 500\,$\upmu$s pulse of 7\,$\upmu$W orange (594\,nm) light~\cite{shields_efficient_2015}. We employ optical filters to restrict the collected NV photoluminescence to the band 691-720\,nm, so that only NV$^-$ fluorescence is collected. Photon counts collected under orange illumination thus serve as a measurement of the NV$^-$ population, which has been shown to convert to NV$^0$ upon capture of photogenerated holes~\cite{jayakumar_optical_2016, lozovoi_optical_2021, wood_wavelength_2024}. Monitoring the NV$^-$ charge state then serves to indicate the transport of holes~\cite{wood_3d-mapping_2024}, while measuring the electric field and space charge serves to indicate where photogenerated electrons flow.

Figure \ref{fig:fig3}(e) depicts the NV charge state dynamics under the same experimental conditions as used previously for measurement of space charge dynamics (Fig. \ref{fig:fig3}(d)) but implementing optical NV charge state detection. With equivalent timescales between the two measurements, we observe that the space charge grows rapidly compared to NV$^-$ hole capture, which only becomes apparent after generation pulses with $\tau_G > 20\,\upmu$s. More significant however is the imputed direction of hole transport: the bias to the right in Fig. \ref{fig:fig3}(e) for NV$^-$ hole capture in fact corresponds to holes generated by beam-2 moving towards the positive electrode, that is, against the direction of the applied electric field (as shown in the inset). 

This seemingly counter-intuitive observation is readily explained by electrometry, which show that the space charge field first cancels the applied electric field to the left of beam-2 (consistent with the direction of electron transport) before equilibrating nearly symmetrically. While the direction of hole drift is set by the total electric field, the hole capture probability of NV$^-$ has been found to depend on the absolute strength the total field as well, with $|\mathbf{E}|\approx 0$ yielding the highest hole capture cross section ~\cite{lozovoi_detection_2023}. Thus, significant NV$^-$ hole capture first appears where the total electric field strength is zero, a region set by the time-dependent space charge formation. 

{\bf Discussion.} The large extent over which $\mathbf{E}_\text{app}+\mathbf{E}_\text{sc} = 0$ reveals interesting clues as to the structure of the space charge regions. Specifically, any model where the space charge field is generated by a nanoscale, polarised configuration of charge traps in the immediate vicinity of each NV centre is difficult to reconcile with our charge transport results, which show that carrier transport is responsible for spreading the space charge region over areas significantly greater than both the local charge neighbourhood of each NV charge ($\sim$nm) and that of the laser illumination spot ($1\,\upmu$m). We are thus left to conclude that the space charge field is formed from a macroscopic configuration of widely separated charges on the scale of tens of microns for the timescales we examine. The charge state of an agglomeration of defects some 10-20\,$\upmu$m away from a given NV centre has a measurable effect on its spin state, and a dominant effect on its local charge dynamics. 

The immediate implications of our work are for diamond-based electrometry and photoelectric detection. For electrometry, the already low instrinsic sensitivity to electric fields of the NV ground state is further reduced by the formation of screening space-charge fields. In particular, high-sensitivity electrometry with dense ensembles of defects (and concomitantly higher nitrogen densities) is likely to be strongly affected by space charge screening~\cite{chen_high-sensitivity_2017}. The linear power dependence of space charge formation we have established in this work has the added implication that reduced optical intensity will not ameliorate screening, as would be the case if NV charge cycling ($\propto I^2$) were the causative factor. For photoelectric detection, the formation of a space charge, and indeed carrier trapping by NV defects, is evidence of reduced photocurrent yield. The presence of space charge screening invalidates the central experimental prerequisite, that an electric field sweeps charge carriers towards the collection electrodes. Space charge effects screen this applied field, meaning charge transport is isotropic about the illumination point. As we have also shown in this work, the probability of carrier capture is dependent on space charge screening, leading to a complicated interplay of space charge formation, propagation and passivation which appears to occur over significant length scales.

From a different perspective, the results of this work open up many avenues of future inquiry. We have shown that significant electric fields -- of order 10-40\,kV\,cm$^{-1}$ -- can essentially be written and stored into the diamond substrate. While measurement of the lifetime of these electric fields is reserved for future work, there is a strong likelihood that the spatial configuration of nitrogen charges responsible for space charge screening will be as stable as optically-addressable defects in diamond, such as the NV centre~\cite{dhomkar_long-term_2016, monge_reversible_2023} or SiV centre~\cite{wood_room-temperature_2023}. Given the typical long-term stability of deep-level charge states in high-purity diamonds such as that used in this work, these `giant' space charge fields are expected to be as stable in the dark as NV charge states, which persist beyond hours or even days in well-defined spatial configurations. Further work could also investigate what effects the built-in electric field from a polarised charge environment has on spectral diffusion~\cite{ji_correlated_2024, delord_correlated_2024}, charge stability~\cite{rieger_fast_2024, bathen_electrical_2019, zuber_shallow_2023}, photoionisation rates~\cite{hanlon_enhancement_2023}, spin coherence~\cite{zheng_coherence_2022, wang_manipulating_2023}, or in turn examine how NV-P1 spin dynamics are affected by charge polarisation~\cite{goldblatt_sensing_2024}.

In conclusion, we have used electrometry of quantum spin defects inside a wide-bandgap semiconductor to probe the formation of space charges, which accumulate due to carrier photogeneration and trapping in the presence of an applied electric field. Our measurements implicate substitutional nitrogen as the principal charge trap that mediates space charge formation, and have shown that charge transport drives the growth of space charges to tens of microns in extent despite tightly focused illumination in a $<$1\,$\upmu$m spot. Within these space charge regions, the applied electric field and space charge field sum to zero, a situation that leads to reduced sensitivity to electric fields and unbiased charge transport in photoelectric detection.                

\section*{Acknowledgments}
We thank C. A. Meriles for stimulating discussions. This work was supported by the Australian Research Council (DE210101093). R. M. G. was supported by an Australian Government Research Training Program (RTP) Scholarship.

\end{document}